%% file: rdpaper.tex
\pdfoutput=1
\documentclass[epjc3,5p,times,twocolumn,a4paper,showpacs,showkeys,amsmath,amssymb,superscriptaddress,nofootinbib]{svjour3}
  
\usepackage[pdftex]{graphicx}
\DeclareGraphicsExtensions{.pdf,.jpeg,.png,.eps}
\usepackage{amssymb}
\usepackage{amsmath}
\usepackage{txfonts}
\usepackage[british]{babel}
\usepackage{url} 
\usepackage{hyperref}

\journalname{Eur. Phys. J. C}

\newcommand*{\BR}     {{\cal B}}
\newcommand*{\meg}        {\mathrm{\mu} \to \mathrm{e} \mathrm{\gamma}}
\newcommand*{\megsign}  {\mathrm{\mu}^+ \to \mathrm{e}^+ \mathrm{\gamma}}
\newcommand*{\meeesign}        {\mathrm{\mu}^+ \to \mathrm{e}^+\mathrm{e}^-\mathrm{e}^+}

\newcommand*{\rmd}        {\mathrm{\mu} \to \mathrm{e} \mathrm{\nu}\bar{\mathrm{\nu}}\mathrm{\gamma}}
\newcommand*{\rmdsign}        {\mathrm{\mu}^+ \to \mathrm{e}^+ \mathrm{\nu}\bar{\mathrm{\nu}}\mathrm{\gamma}}
\newcommand*{\michel}        {\mathrm{\mu} \to \mathrm{e} \mathrm{\nu}\mathrm{\bar{\nu}}}

\newcommand*{\egamma}      {E_{\mathrm{\gamma}}}
\newcommand*{\tgamma}       {t_{\mathrm{\gamma}}}
\newcommand*{\epositron}      {E_\mathrm{e}}
\newcommand*{\tpositron}      {t_\mathrm{e}}
\newcommand*{\tegamma}        {t_\mathrm{{e\gamma}}}

\newcommand*{\thetaegamma}    {\theta_\mathrm{{e\gamma}}}
\newcommand*{\phiegamma}      {\phi_\mathrm{{e\gamma}}}
\newcommand*{\thetae}    {\theta_\mathrm{e}}
\newcommand*{\phie}      {\phi_\mathrm{e}}
\newcommand*{\thetagamma}    {\theta_\mathrm{\gamma}}
\newcommand*{\phigamma}      {\phi_\mathrm{\gamma}}
\def\vector#1{\mbox{\boldmath $#1$}} 
\begin{document}
\title{
Measurement of the radiative decay of polarized muons in the MEG experiment 
}
\subtitle{The MEG Collaboration}
\date{Received: 21 Dec 2015 / Accepted: 9 Feb 2016}
\include{author_INFNComb-epjc}

\thankstext[*]{e1}{Corresponding author: uchiyama@icepp.s.u-tokyo.ac.jp}
\thankstext[$\S$]{section}{Presently at INFN, Laboratori Nazionali di Frascati, Via E.~Fermi, 40-00044 Frascati (Roma), Italy.}
\thankstext[$\dag$]{dagger}{Deceased.}
\maketitle

\begin{abstract}
We studied the radiative muon decay $\rmdsign$ by using for the first time an almost fully polarized muon source. 
We identified a large sample ($\sim\!13000$) of these decays in a total sample of $1.8\times10^{14}$ positive muon decays collected in the MEG experiment in the years 2009--2010
and measured the branching ratio $\mathcal{B}(\rmd) = (6.03\pm0.14\mathrm {(stat.)}\pm0.53\mathrm{(sys.)})\times 10^{-8}$ for $\epositron>45~\mathrm{MeV}$ and $\egamma>40~\mathrm{MeV}$, consistent with the Standard Model prediction. 
The precise measurement of this decay mode provides a basic tool for the timing calibration, a normalization channel, and a strong quality check of the complete MEG experiment in the search for $\megsign$ process. 
\end{abstract}

\section{Introduction}\label{sec:intro}
In the Standard Model of particle physics (SM), muons decay through the purely leptonic weak interaction: the tree level process is $\michel$ (Michel decay). 
This decay has been carefully studied since the discovery of the muon 
and still provides one of the most useful tools for studying the weak interactions. 
Radiative muon decay, $\rmd$ (RMD), is the first order QED correction to Michel decay with the additional emission of one inner bremsstrahlung photon.

The importance of studying RMD is twofold: on one hand, it provides a tool for
investigating weak interactions 
since it is sensitive to some of parameters appearing in the most general formula of muon decay; this approach is followed in e.g.
\cite{Eichenberger84,munyangabe_2012}.
On the other hand, it constitutes important sources of background for experiments searching for rare muon decays, not allowed in the minimal SM, such as $\megsign$.
RMD events form a time-correlated background for the $\megsign$ search when the two neutrinos carry away so little momentum that the RMD event falls within the signal window for $\megsign$ events, determined by the experimental resolutions. Moreover, high energy $\gamma$-rays from RMD events constitute 
the dominant accidental background for experiments operating at high muon stopping rates, by random time-overlapping with high energy positrons from Michel decays.
Finally, the identification of the time-correlated peak due to RMD events allows a calibration of the positron--photon relative timing as well as a measure of the associated resolution and provides a strong internal consistency check for the $\megsign$ analysis.

\section{MEG experiment}
\label{sec:meg}
The MEG experiment has been searching for the $\megsign$ decay since 2008  \cite{meg2009,meg2010} at Paul Scherrer Institut in Switzerland \cite{PSI} reaching the most stringent upper limit up to date on the $\megsign$ branching ratio based on the data sample collected in 2009--2011 \cite{meg2013}. 
The experiment is briefly described below; a full description is available in \cite{megdet}. 

In this paper, we use a cylindrical coordinate system ${\it (r,\phi,z)}$ with origin at the centre of MEG and the $z$-axis being parallel to the incoming muon beam.
Where used, the polar angle $\theta$ is defined with respect to the $z$-axis.

A high intensity positive muon
beam is brought to rest in a 205~$\mathrm{\mu m}$ thick slanted plastic target, placed
at the centre of the experimental set-up. 
MEG uses surface muons, originating from pion decays at rest, at the surface of the production target. Hence, they are fully polarized at their origin.  The depolarization mechanisms along the beam-line and in the stopping target have been estimated in detail and are small and under control. 
The residual muon polarization at the decay point along the beam axis is measured to be \cite{polarization}
\begin{equation}
P_{\mu^+} = -0.85 \pm 0.03 ~ \mathrm{(stat.)} ~ ^{+ 0.04}_{-0.05} ~ \mathrm{(sys.)}
\label{eq:pol}
\end{equation}
from the angular distribution of decay positrons, in agreement with expectations.

The muon decay products are detected by a liquid xenon (LXe) photon detector and a positron spectrometer with a gradient magnetic field generated by the superconducting magnet COBRA.
The LXe detector consists of 900~$\ell$ LXe and 846 photomultiplier tubes and measures energy, interaction time and position of the photon.
Its geometrical acceptance  is $\thetagamma \in (70^{\circ},110^{\circ})$ and $\phigamma \in (-60^{\circ},60^{\circ})$ covering $\sim\!11$\% of the total solid angle. 
The opposite angular region is covered by the spectrometer consisting of a set of 16 drift chambers, radially aligned, for the measurement of the positron momentum, complemented by a timing counter (TC), composed of two scintillator arrays, for the measurement of the positron timing.
The MEG detector and the trigger are optimized to search for $\megsign$ events. Therefore, there is only a limited energy and angular window to detect RMD events.

The time ($\tpositron$) and vertex of the positron at the target are obtained by extrapolating the time measurement at the TC back along the track trajectory. 
The photon time ($\tgamma$) is calculated by connecting the photon interaction position in the LXe volume to the positron vertex on the target and extrapolating the time measurement at the LXe detector back to the target.

The kinematics of the events is described by five observables: the photon and positron energies ($\egamma$, $\epositron$), their relative directions ($\thetaegamma$, $\phiegamma$),\footnote{$\thetaegamma = (\pi-\thetae) - \thetagamma$ and
$\phiegamma = (\pi + \phie) - \phigamma$.}
and the emission time ($\tegamma = \tgamma - \tpositron$).

A dedicated trigger system allows an efficient pre-se\-lec\-tion of $\megsign$ candidate events (the MEG trigger), with an almost zero dead-time \cite{trigger2013,Galli:2014uga}. Background is efficiently suppressed by an on-line requirement of a positron and a photon close to their kinematic limit moving in opposite direction in time coincidence.
In parallel to the main trigger, several other triggers are activated in a physics run.
In this analysis, we select RMD events from the MEG trigger data  while other trigger data are used for the calibration of the detectors and the normalization.

Several kinds of dedicated runs are frequently taken at different intervals to calibrate and monitor the detectors.
Among them a run to calibrate the LXe detector with high energy photons close to the signal region is especially important. In this run, neutral pions are produced through the charge-exchange reaction
$\pi^{-} \mathrm{p} \rightarrow \pi^{0} \mathrm{n}$, by using a negative pion beam
brought to rest in a liquid-hydrogen target. 
The photons from neutral pion decay $\pi^0\rightarrow \gamma \gamma$ are used to calibrate the LXe detector. 
A counter formed by nine NaI(Tl) crystals\footnote{In 2011 this detector was replaced by a higher resolution BGO array.} is placed on the opposite side of the LXe detector to tag one of the $\gamma$-rays, yielding an almost monochromatic source of 55 and 83~MeV photons. 
The photon energy, timing and position resolutions as well as the energy scale are measured in this run.

\begin{figure}[tbp]
\centering
\includegraphics[width=18pc]{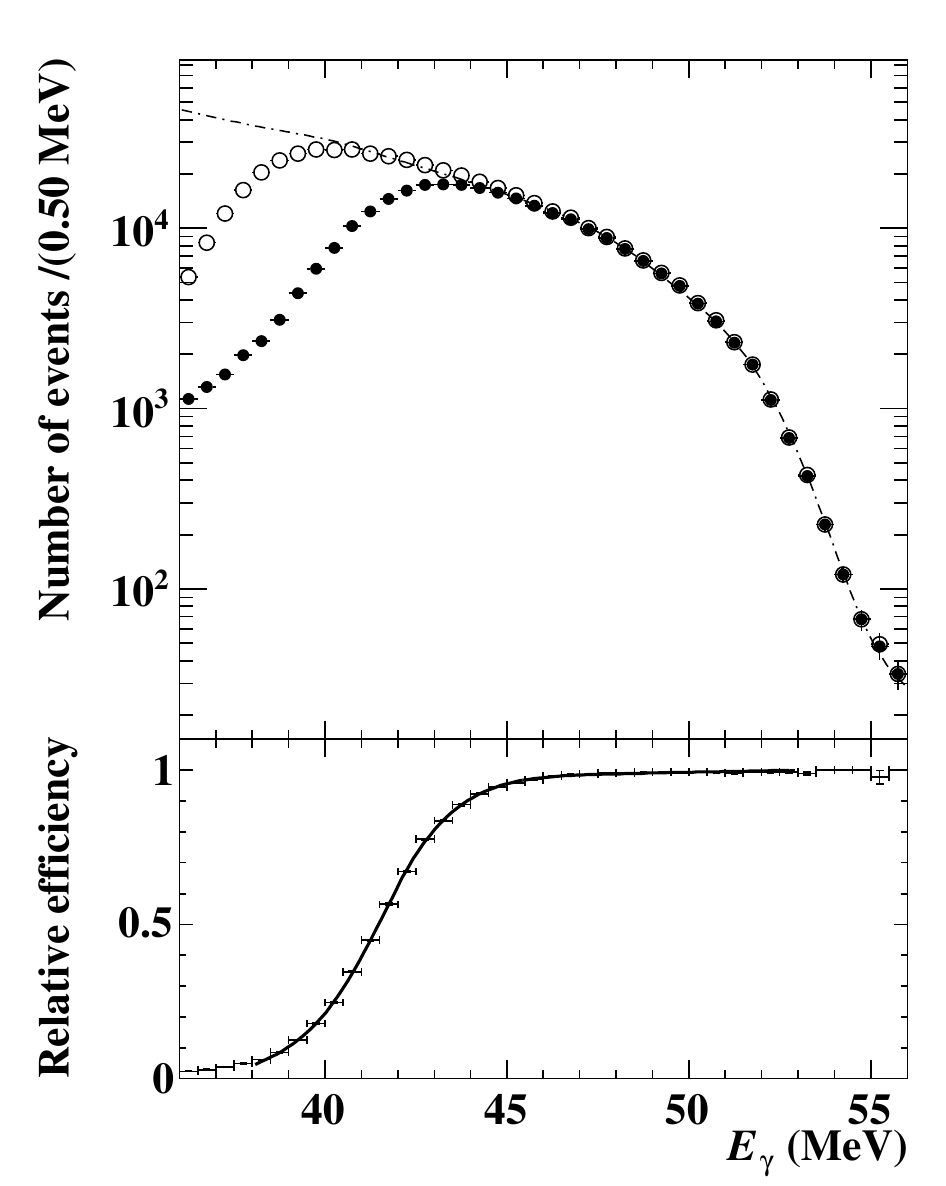}
\caption{\label{fig:GammaEfficiency}Photon energy spectra with different trigger
thresholds (top). The solid (open) circles correspond to the normal (lower) threshold. The dot-dashed line is the MC simulation spectrum smeared with the
detector response; the calculated spectrum is used to correct for the trigger
effect in the lower threshold distribution.
The bottom plot shows the ratio of the normal threshold spectrum to the lower threshold one.}
\end{figure}

The photon energy is limited by the trigger threshold, $\egamma \gtrsim 40~\mathrm{MeV}$.   
A pre-scaled trigger with a lowered $\egamma$ threshold (by $\sim\!4$~MeV) is enabled during the normal physics run. This allows a relative measurement of the energy-dependent efficiency curve of the LXe detector (Fig.~\ref{fig:GammaEfficiency}),
while the absolute photon detection efficiency is evaluated via MC simulation \cite{megdet}.
The position dependence of the detection efficiency is investigated and the average value is calculated by taking into account the observed event distribution.
This evaluation is cross-checked by measuring the probability of detecting one of the
two photons from a neutral pion decay in the LXe detector under the condition that the
other photon is detected by the NaI(Tl) counter.
The measurement and the MC simulation agree to within 2\%; the spread is considered as an estimate of the systematic uncertainty, resulting in the detection efficiency $\epsilon_\gamma=0.63\pm0.02$ at the plateau. 
 
The LXe detector also exhibits good linearity; the non-linearity of the energy scale is found to be $<0.1$\%, estimated from the 55 and 83~MeV photons from the pion decays as well as from the $17.7~\mathrm{MeV}$ peak position of the $^7\mathrm{Li}(\mathrm{p},\gamma)^8\mathrm{Be}$ reaction induced by using a Cockcroft-Walton proton accelerator \cite{calibration_cw}. 
The uncertainty of the energy scale around the signal region is evaluated to be $0.3$\% from combining several kinds of calibration data. 

\begin{figure}[tbp]
\centering
\includegraphics[width=18pc]{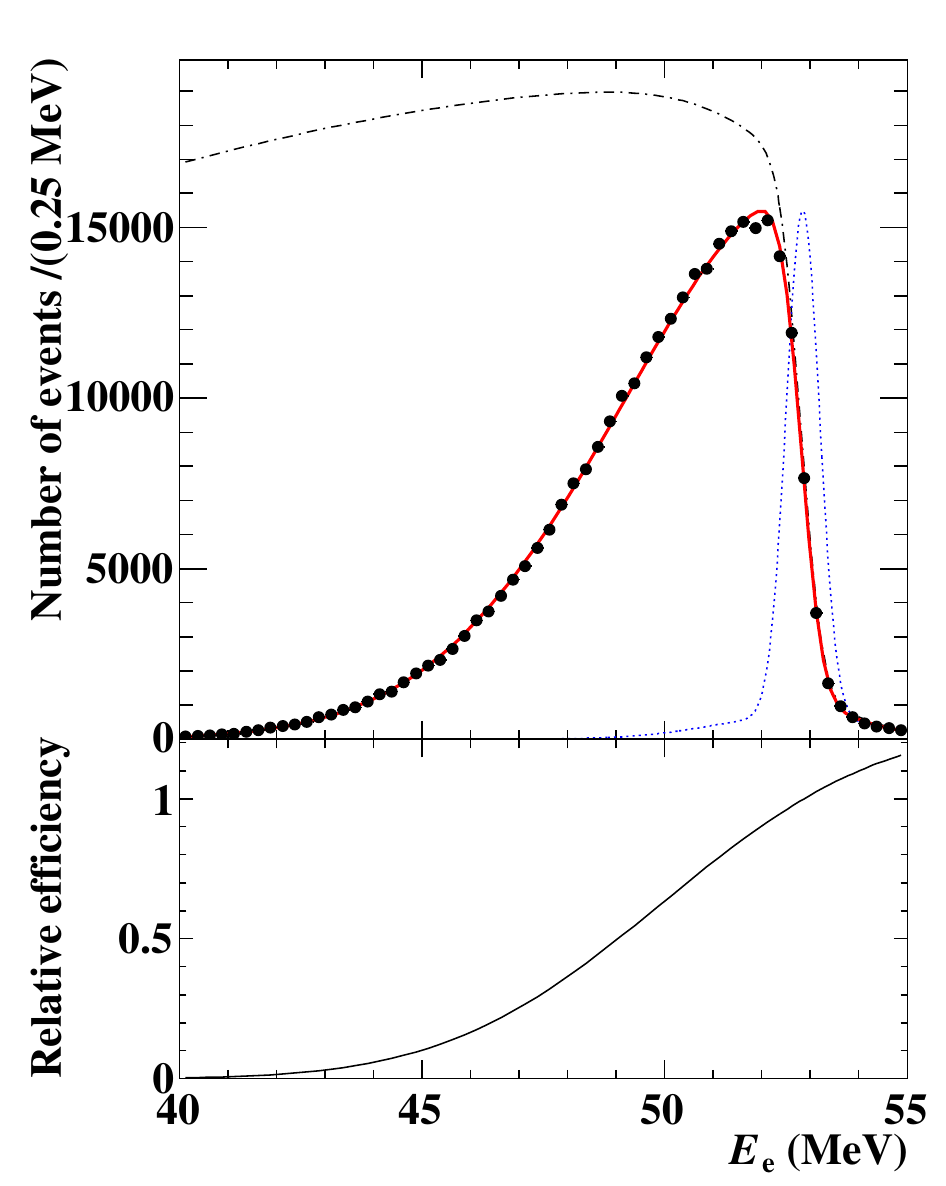}
\caption{\label{fig:MichelFit}Michel positron spectrum (top). The dots are
data and the red-solid line is the best-fit function. The blue-dotted line shows the
detector resolution and the dot-dashed line shows the theoretical Michel spectrum, folded
with the detector resolution. The bottom plot shows the energy dependence of the acceptance
extracted from the fit and normalized to 1 at $52.8$ MeV.}
\end{figure}

The spectrometer preferentially selects high energy positrons,  with $\epositron \gtrsim 45~\mathrm{MeV}$.
The Michel positron spectrum is used as a calibration tool for the spectrometer by comparing the measured one with the precisely known theoretical one, including the first order radiative corrections \cite{kinoshita_1959}. 
The resolution and the energy-dependent efficiency 
are simultaneously extracted by fitting the theoretical Michel spectrum folded with the detector response to the measured spectrum, as shown in Fig.~\ref{fig:MichelFit}. 
The absolute positron detection efficiency is not needed because of the normalization scheme adopted (described in Sec.~\ref{sec:br}).

As reported in \cite{meg2010}, the resolutions for positrons and photons with energies 
close to the kinematic limit $m_{\mu}/2$ are $\sigma(\epositron) \sim 330~\mathrm{keV}$,
$\sigma(\egamma) \sim 1.0~\mathrm{MeV}$, $\sigma(\thetaegamma) \sim 17~\mathrm{mrad}$, $\sigma(\phiegamma) \sim 14~\mathrm{mrad}$,
$\sigma(\tegamma) \sim 130~\mathrm{ps}$.

Although measurements of RMD have already been obtained by other experiments \cite{PDBook_2014}, 
the MEG data provides the unprecedented opportunity of measuring RMD from polarized muons at the kinematic edge.

\section{Distribution of radiative muon decay}

The RMD differential branching ratio was calculated by several 
authors \cite{Lenard53,Sirlin56,fronsdal_1959_pr,eckstein_1959_ap}. 
In the framework of the $V-A$ theory of weak interactions, it reads as \cite{kuno_2001} 
\begin{align}
 d\BR(\rmdsign) &=
\frac{\alpha}{64\pi^3} \beta \, dx \,\frac{dy}{y}\, d\Omega_\mathrm{e} \,d\Omega_{\gamma} \,\big[ F(x,y,d) \nonumber\\ 
 & - \beta \vector{P_{\mu^+}} \cdot \vector{\hat{p}}_\mathrm{e} G(x,y,d)
  \nonumber\\
 &  - \vector{P_{\mu^+}} \cdot \vector{\hat{p}}_\gamma H(x,y,d)
  \big],
 \label{eq:RDDifferential} 
\end{align}
where $x=2\epositron/m_\mu$, $y=2\egamma/m_\mu$, 
$\vector{\hat{p}}_k$ is the unit vector of the particle $k$  (positron or photon) momentum in the muon rest frame, 
$\vector{P_{\mu^+}}$ is the muon polarization vector, 
and $d = 1 - \beta \vector{\hat{p}}_e \cdot  \vector{\hat{p}}_\gamma$. 
Detailed descriptions of the functions $F$, $G$ and $H$ are given in \cite{kuno_2001}.
A few authors calculated the higher order corrections for some special cases \cite{Fischer:1994pn,Arbuzov:1998kr,Arbuzov:2004wr} 
and only recently a full next-to-leading order (NLO) calculation became available \cite{Fael:2015}.
In this paper only the lowest-order general calculation is used.

\begin{figure}[tb]
\centering
\includegraphics[width=18pc]{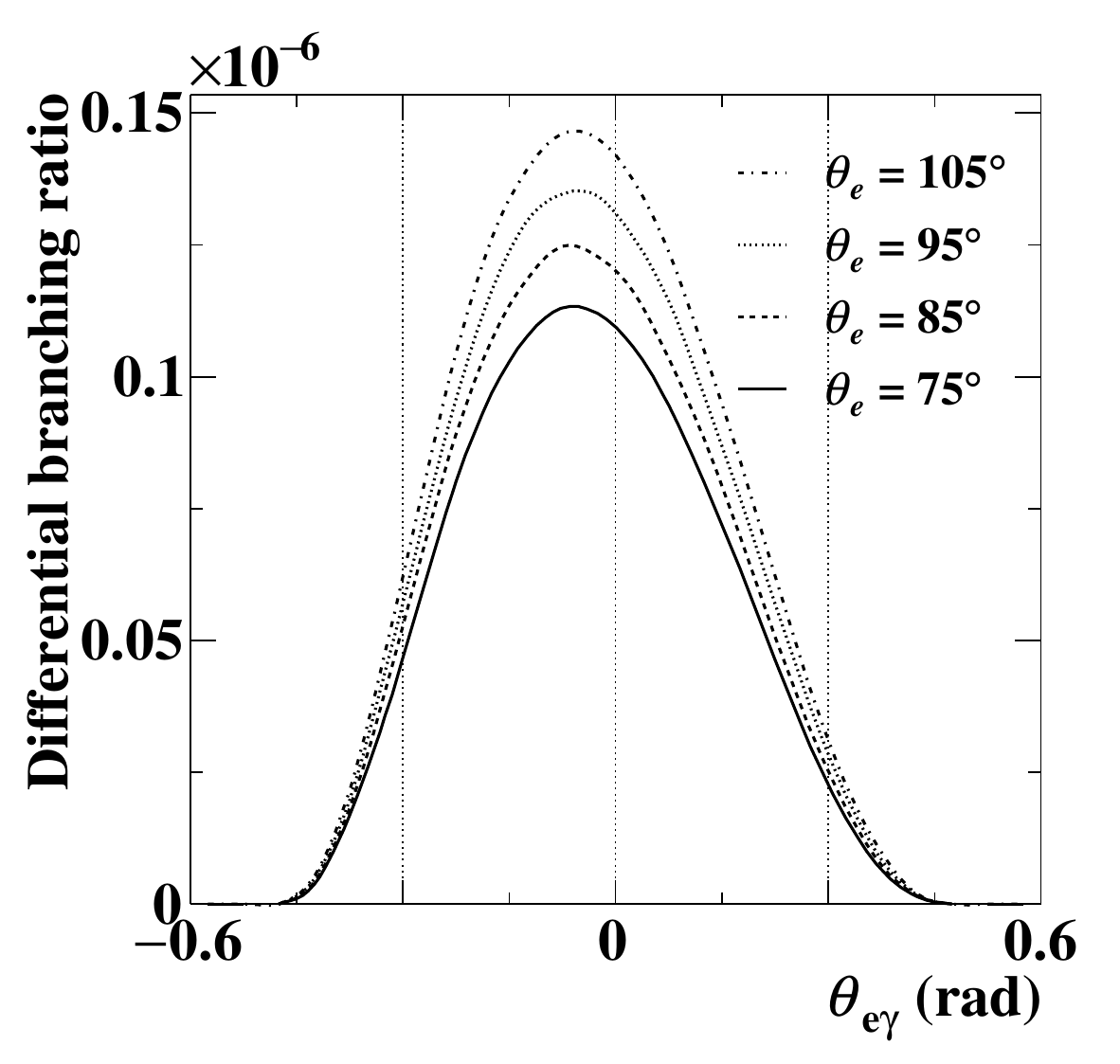}
\caption{\label{fig:RMDTheta}Differential branching ratio of RMD for $P_{\mu^+}=-0.85$ as 
a function of $\theta_{\mathrm{e}\gamma}$ for four different values of positron polar angle $\thetae$. These distributions are obtained by 
the numerical integration of Eq.~(\ref{eq:RDDifferential}) over  $\epositron>45$, $\egamma>40$~MeV, and $\phiegamma$.}
\end{figure}

The relative angle distribution shows an asymmetric shape in $\thetaegamma$, while the distribution in $\phiegamma$ remains symmetric. 
The $\thetaegamma$ distributions for polarized muons with $P_{\mu^+}=-0.85$ for four different values of $\thetae$ are shown in Fig.~\ref{fig:RMDTheta} after integration over $\phiegamma$ and the positron and photon energies. 
The relative-angle range kinematically allowed\footnote{from energy and momentum conservation} for RMD is so restricted by the energy selection imposed on the positron and photon that it is fully covered by the MEG detector and trigger.
However, the distribution is somewhat distorted due to the 
energy-dependent variation of the trigger efficiency over the angular range, as explained below.

\begin{figure}[tbp]
\centering
  \includegraphics[width=17pc] {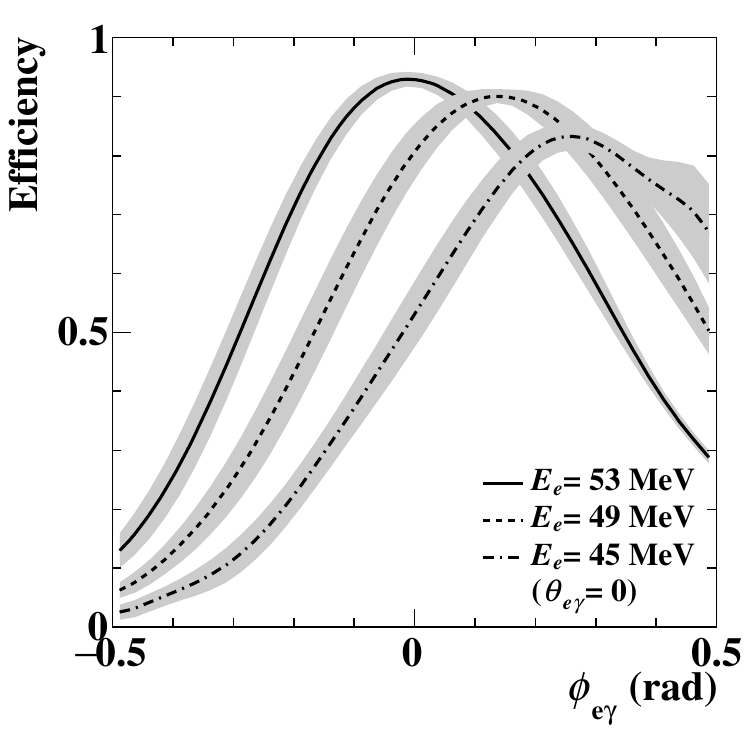} \\
  \includegraphics[width=17pc] {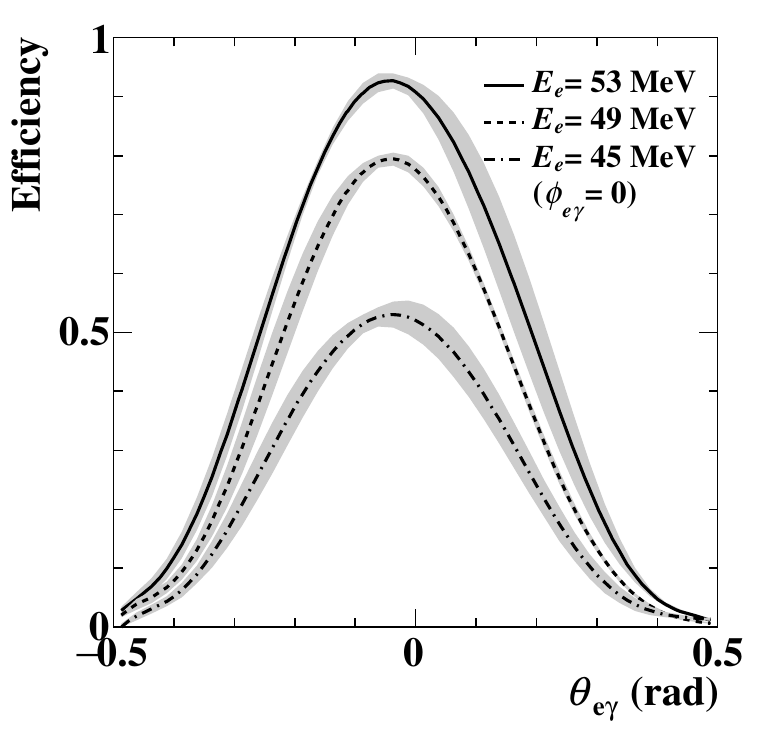}
 \caption[]{\label{fig:TRGEfficiency}Efficiencies of the trigger directional
 match selection versus relative angles. The 
bands around the curves show the uncertainties ($1\sigma$).}
\end{figure}

The directional match efficiency of the trigger is evaluated via MC
simulation and the distribution of the accidental background. 
Because the spectrometer response introduces a correlation
in the distribution of the positron emission angle and momentum, 
the relative angle
distribution, after the directional match selection induced by the $\megsign$ trigger,
is asymmetric and dependent on the positron energy. Therefore, the directional match
efficiency is calculated for different values of $\epositron$, as shown in
Fig.~\ref{fig:TRGEfficiency}. The spread between calculations and
measurements is considered as an estimate of the systematic uncertainty.
%

\section{Measurement of radiative muon decay}

The data sample used in this analysis corresponds to $\sim\! 1.8\times10^{14}$ positive muon decays
in the target, collected in 2009--2010.\footnote{MEG ended its run in 2013. This sample corresponds to about one fourth of the full MEG data-set.} We used events reconstructed in the analysis
window defined as $45<\epositron<53$~MeV, $40<\egamma<53~\mathrm{MeV}$, $|\phiegamma|<0.3$~rad
and $|\thetaegamma|<0.3$~rad. 
The event reconstruction and event selection as well as the data sample for this study are identical to those for the $\megsign$ search in \cite{meg2010}.
A complete description of the MEG analysis procedure is given in \cite{megdet,meg2009,meg2010}.

The main background to the RMD signal comes from the accidental coincidence of positrons and
photons originating from different muon decays. 
Because the two particles are
uncorrelated, the accidental background events are distributed randomly with respect to $\tegamma$. 
On the other hand, the positron and the photon from a RMD are emitted simultaneously; therefore, the presence of RMD events is signalled by a peak around zero in the $\tegamma$ distribution and it is well described by a sum of two Gaussian functions.\footnote{The broad component is mainly due to multiple Coulomb scattering of the positron in material placed between the drift chamber active volume and the TC (support frame, preamplifiers, and cables of the drift chambers), resulting in a worse extrapolation of the positron trajectory between them and hence in a larger error in the time-of-flight calculation.}

\begin{figure}[tb]
\centering
\includegraphics[width=17pc] {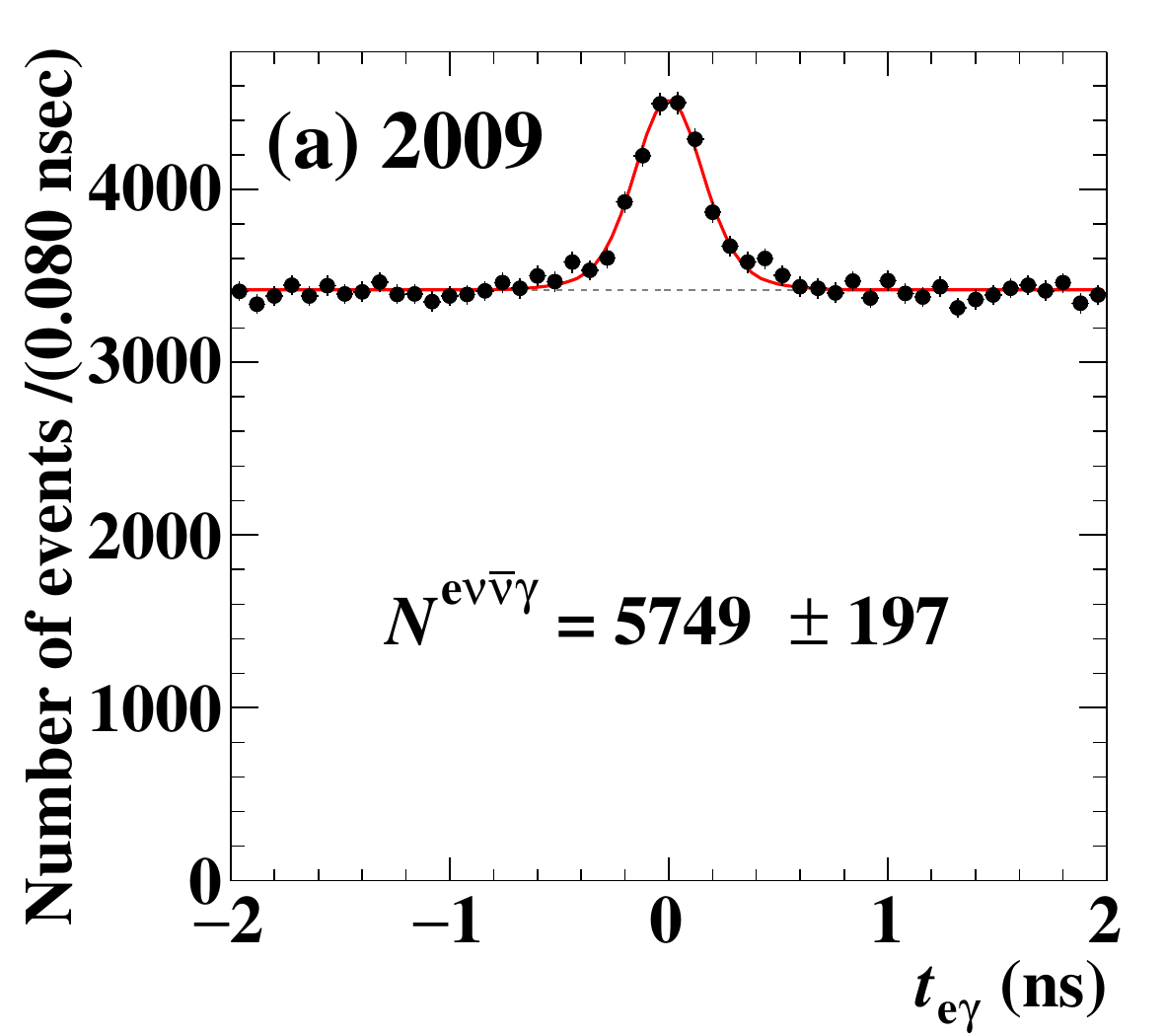} \\
\includegraphics[width=17pc] {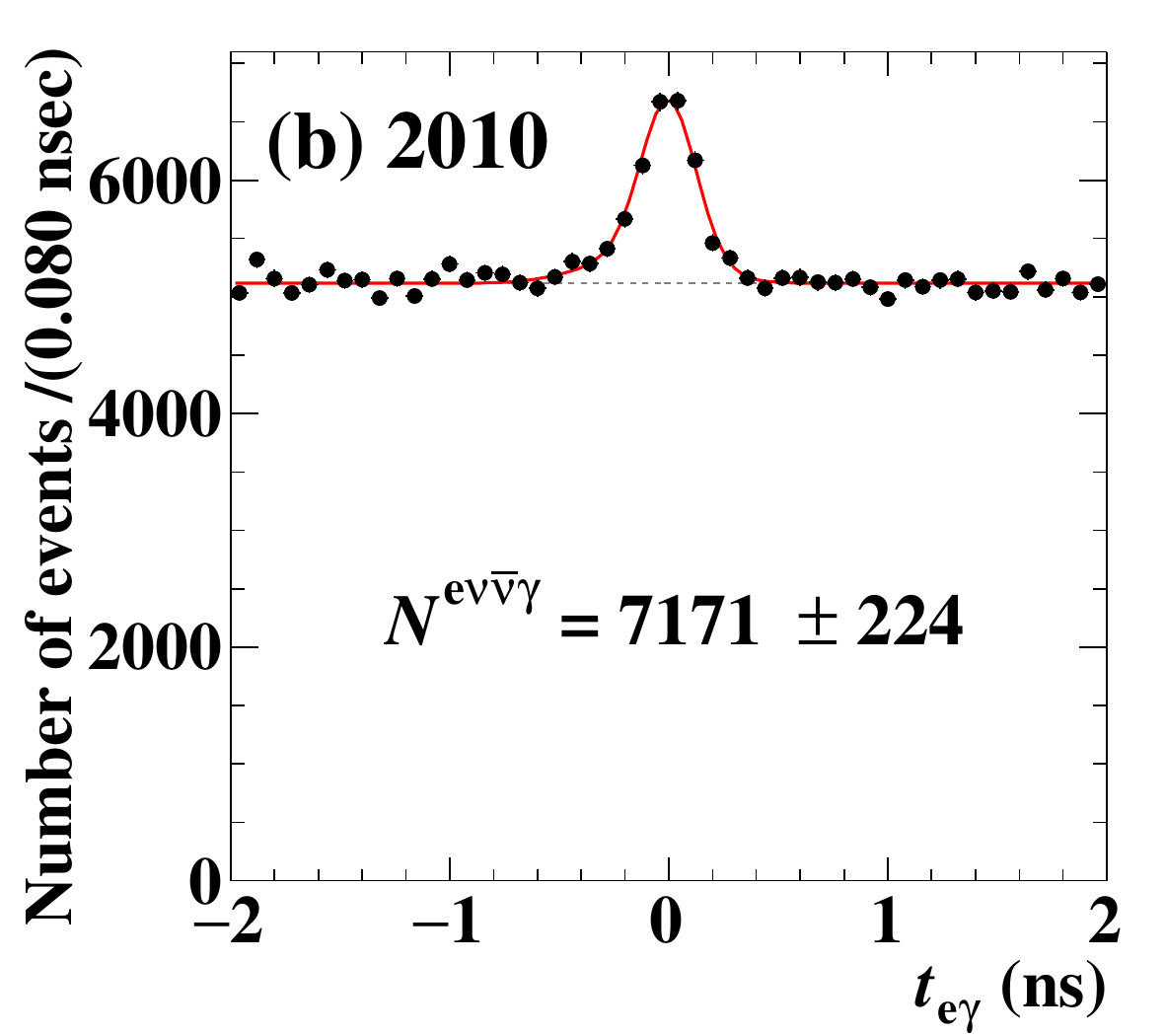}
\caption{\label{fig:RMDTimeFit}Distributions of $t_{e\gamma}$ in (a) 2009 data and (b) 
2010 data. The best-fit functions of the sum of the RMD and the accidental-background PDFs 
(red solid) and those of the accidental-background only (dashed) are superimposed. }
\end{figure}

\begin{figure*}[htb]
\centering
\includegraphics[width=.9\textwidth]{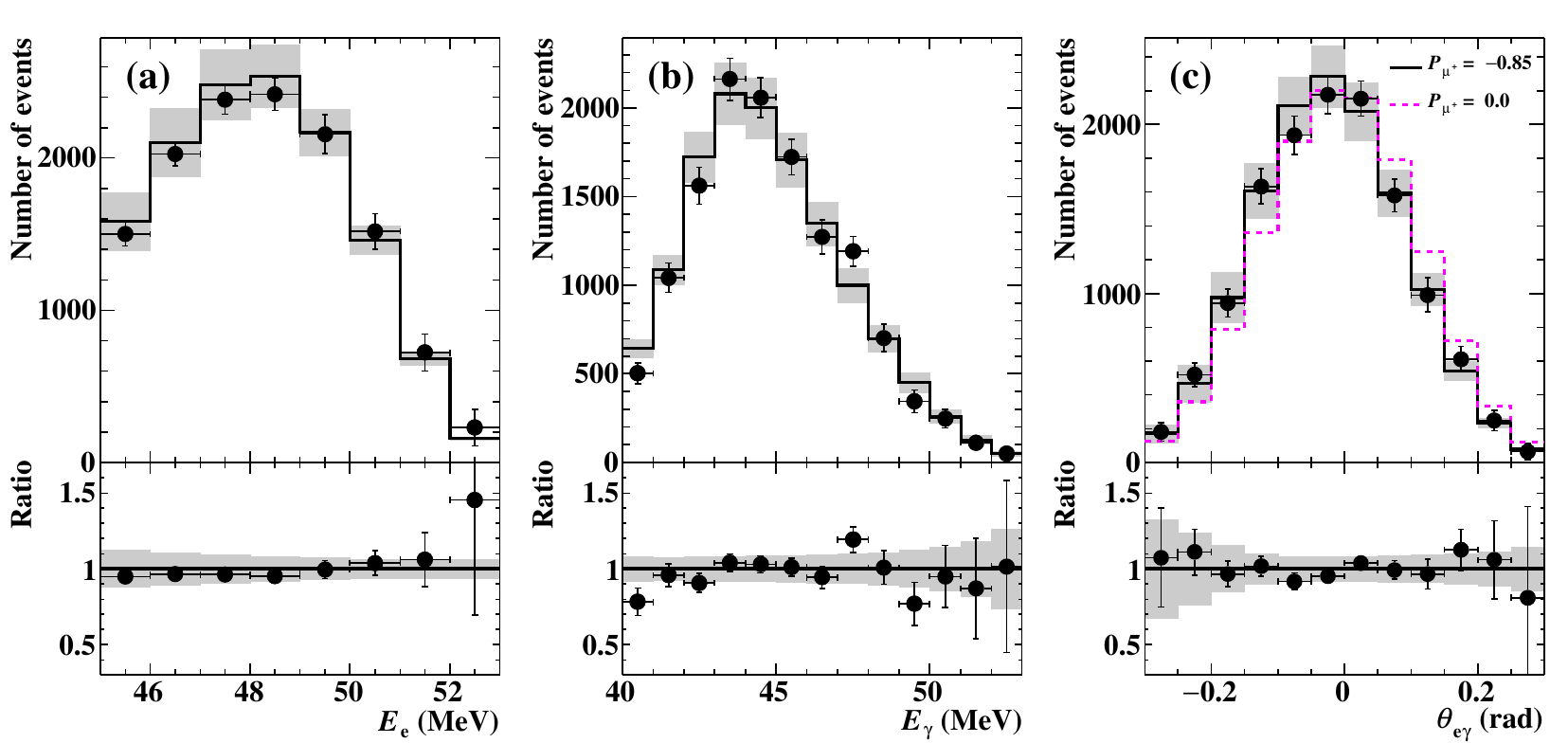}
\caption{\label{fig:RMDProjectedDistribution}Projected distributions of RMD events in
(a) $E_\mathrm{e}$, (b) $\egamma$ and (c) $\theta_{\mathrm{e}\gamma}$.
The solid circles are measurements obtained from the 2009 and 2010 data sets, the histograms show the expectations assuming
$P_{\mu^+} = -0.85$ and the normalization based on Michel positron events. The bands
show the systematic uncertainties in the calculation of the expectations. The bottom plots
show the ratio of the measurements to the expectations. In (c) the expected
distribution calculated assuming $P_{\mu^+} = 0$ (magenta dashed line) is also superimposed for comparison.}
\end{figure*}

To measure the number of RMD events ($N^{\mathrm{e}\nu\bar{\nu}\gamma}$), we fitted a probability density function (PDF), given by the sum of the RMD PDF and the accidental background PDF (a uniform
distribution), to the $t_{\mathrm{e}\gamma}$ distribution (Fig.~\ref{fig:RMDTimeFit}). 
We separately analysed $2009$ and $2010$ data because of the different time resolutions: the electronics were improved in the time measurement of 2010.
To measure the distribution of RMD in terms of energy and angle, the fits were repeated
for data-sets divided into bins. 
Figure~\ref{fig:RMDProjectedDistribution} shows the experimental distributions of RMD events in $\epositron$, $\egamma$ and $\thetaegamma$. 
%

\section{Results}

\subsection{Branching ratio measurement}\label{sec:br}

Since the total branching ratio for RMD is infrared divergent,
a well defined measure of the branching ratio requires a region of the phase space which includes a lower limit on the photon energy.
Here we measure the branching ratio for the largest phase space allowed by our detector set-up, namely for $\epositron>45$ and $\egamma>40~\mathrm{MeV}$.

To convert the number of measured RMD events into the branching ratio, it is normalized to the number of Michel positrons counted simultaneously. This is accomplished by a pre-scaled unbiased Michel positron trigger:
\begin{align}
N_\mu = \frac{N^{\mathrm{e}\nu\bar{\nu}}}{f^{\mathrm{e}\nu\bar{\nu}}_{E_\mathrm{e}}}\times 
\frac{p^{\mathrm{e}\nu\bar{\nu}}}{\epsilon^{\mathrm{e}\nu\bar{\nu}}_{\mathrm{trg}}} 
 \times \frac{1}{\langle\epsilon^{\mathrm{e}\nu\bar{\nu}}_{\mathrm{e}}\rangle},
\end{align}
where $N^{\mathrm{e}\nu\bar{\nu}}$ is the number of detected Michel positrons, $f^{\mathrm{e}\nu\bar{\nu}}_{E_\mathrm{e}}$ is the fraction of Michel spectrum in the corresponding energy range, and 
$p^{\mathrm{e}\nu\bar{\nu}}=10^7$ is the pre-scaling factor of the Michel positron trigger corrected by $\epsilon^{\mathrm{e}\nu\bar{\nu}}_{\mathrm{trg}}=0.88\pm0.01$ to account for the dead time of the trigger scaler.
The positron detection efficiency is $E_\mathrm{e}$ dependent, $\epsilon_\mathrm{e}(E_\mathrm{e}$), as described in Sec.~\ref{sec:meg}, and $\langle\epsilon^{\mathrm{e}\nu\bar{\nu}}_{\mathrm{e}}\rangle$ 
is the weighted average efficiency over the corresponding range of the Michel spectrum.

The branching ratio is calculated as follows,
\begin{align}
&\mathcal{B}(\rmd) = \frac{N^{\mathrm{e}\nu\bar{\nu}\gamma}}{N_\mu \times \langle \epsilon^{\mathrm{e}\nu\bar{\nu}\gamma}\rangle}\nonumber\\	
&= N^{\mathrm{e}\nu\bar{\nu}\gamma}\times
\Biggl(
\frac{f^{\mathrm{e}\nu\bar{\nu}}_{E_\mathrm{e}}}{N^{\mathrm{e}\nu\bar{\nu}}} \times
\frac{\epsilon^{\mathrm{e}\nu\bar{\nu}}_{\mathrm{trg}}}{p^{\mathrm{e}\nu\bar{\nu}}}
\Biggr)
  \times
\frac{\langle\epsilon^{\mathrm{e}\nu\bar{\nu}}_{\mathrm{e}}\rangle}
		{\langle\epsilon^{\mathrm{e}\nu\bar{\nu}\gamma}_{\mathrm{e}}\rangle} \times 
\frac{1}{\langle\epsilon^{\mathrm{e}\nu\bar{\nu}\gamma}_{\gamma}\rangle} \times
\frac{1}{\langle\epsilon^{\mathrm{e}\nu\bar{\nu}\gamma}_{\mathrm{trg}}\rangle}, 
\label{norm}
\end{align}
where $\langle\epsilon^{\mathrm{e}\nu\bar{\nu}\gamma}_{\mathrm{e}}\rangle$, $\langle\epsilon^{\mathrm{e}\nu\bar{\nu}\gamma}_{\gamma}\rangle$, and $\langle\epsilon^{\mathrm{e}\nu\bar{\nu}\gamma}_{\mathrm{trg}}\rangle$ are the weighted average efficiencies for the positron detection, the photon detection, and the trigger directional match, respectively, over the RMD spectrum.
The positron detection efficiencies for the two channels appear in ratio and thus the branching-ratio measurement is insensitive to the absolute value of the positron detection efficiency and independent of the instantaneous beam rate.

The total number of RMD events $N^{\mathrm{e}\nu\bar{\nu}\gamma} =12\,920\pm299$ corresponds to 
\begin{align}
\mathcal{B}(\rmd) = (6.03\pm0.14\pm0.53)\times 10^{-8} \nonumber\\ 
 \mathrm{for} \quad (\epositron>45, \egamma>40~\mathrm{MeV}),
\label{br}
\end{align}
where the first uncertainty is statistical and the second one is systematic.
This result is in good agreement with the SM value calculated by a numerical integration of the theoretical formula (\ref{eq:RDDifferential}), 
$\mathcal{B}^{\mathrm{SM}}(\rmd) = 6.15\times 10^{-8}$ (this estimation does not include the contributions from radiative corrections, see Sec.~\ref{sec:future}).

The overall detection efficiency of RMD events in this region is $\sim\!0.1$\%.
This low efficiency is due to the small geometrical acceptance ($\sim\!10$\%) and the detector and trigger optimization for the detection of $\megsign$ events.

The systematic uncertainties are summarised in Table~\ref{tab:BRSystematic}.
The largest contribution comes from the energy dependence of the positron detection efficiency.
This is due to the correlation between the
acceptance curve and the response function, which are simultaneously extracted in the
Michel spectrum fit, and to the dependence of the positron energy threshold and spectral shape on the positron azimuthal emission angle induced by the directional match of the trigger.
This dependence affects the determination of the normalization factor based on Michel decays, since for events involving isolated positrons the directional match is clearly not imposed; thus, acceptance factors are different between RMD and Michel events and do not cancel out perfectly.

\begin{table}	
 \caption{\label{tab:BRSystematic}Summary of relative uncertainties in the branching ratio measurement.}
\begin{tabular*}{\columnwidth}{@{\extracolsep{\fill}}lc@{}}
    \hline\hline
	Source & (\%) \\\hline
	Photon energy scale 		& 3.4\\
	Photon response \& efficiency curve 	& 2.1\\
	Positron response \& efficiency curve	& 6.1\\
	Time response				& 0.5\\
	Angle response			& $  \mkern-18mu <0.1$\\
	Directional match efficiency	& 1.2\\
	Angle dependence of efficiency	& 0.6\\
	Muon polarization				& $  \mkern-18mu <0.1$\\
	Absolute photon efficiency	& 3.7\\
	Absolute trigger efficiency		& 1.0\\
	Michel positron counting		& 2.8\\
    \hline    
	Total systematic			& 8.8\\
	Statistical				& 2.3\\
    \hline    
    Total (added in quadrature)& 9.1 \\
    \hline\hline
   \end{tabular*}
\end{table} 

\subsection{Spectral analysis}
We also performed a $\chi^2$-fit to the measured spectrum with the polarization and the normalization
as floating parameters in order to study the spectral shape in the three-dimensional space $(\epositron,\egamma,\thetaegamma)$. The data sample was divided into $2\times 2\times 6$ bins
in $(\epositron,\egamma,\thetaegamma)$ respectively (24 bins in
total).
The expected number of events for the bin $i$ was calculated as follows:
$N^\mathrm{cal}_i(P_{\mu^+},\alpha) = {\cal B}_i(P_{\mu^+})\cdot \epsilon_i ^{\mathrm{e}\nu\bar{\nu}\gamma}\cdot \alpha N_{\mu}$,
where the partial branching ratio ${\cal B}_i(P_{\mu^+})$ is given by
the SM value, that depends on the muon polarization $P_{\mu^+}$ according to Eq.~(\ref{eq:RDDifferential}); $\epsilon_i^{\mathrm{e}\nu\bar{\nu}\gamma}$ is the efficiency
for RMD events in this bin;
 and $\alpha$ is a normalization scale parameter,
relative to the normalization based on the Michel positron measurement.
Since the systematic uncertainties introduce correlations among the bins,
we built a covariance matrix $V$. The covariance matrix for each source of
systematic uncertainty was evaluated by calculating the deviation of $N^\mathrm{cal}_i$
when the corresponding parameter was varied by one standard deviation.
The total covariance matrix including the statistical uncertainty is the sum
of the covariance matrices for individual uncertainty sources, except for those
related to the absolute scale, that is, the uncertainties in the Michel positron counting 
and the absolute trigger and photon efficiencies. The $\chi^2$ is defined as:
\begin{equation}
\chi^2(P_{\mu^+},\alpha) = \sum_{i,j=1}^{24}(N^{\mathrm{meas}}_i - N^{\mathrm{cal}}_i)(V_{ij})^{-1}(N^{\mathrm{meas}}_j - N^{\mathrm{cal}}_j).
\label{chi2}
\end{equation}

\begin{figure}[htb]
\includegraphics[width=1\columnwidth]{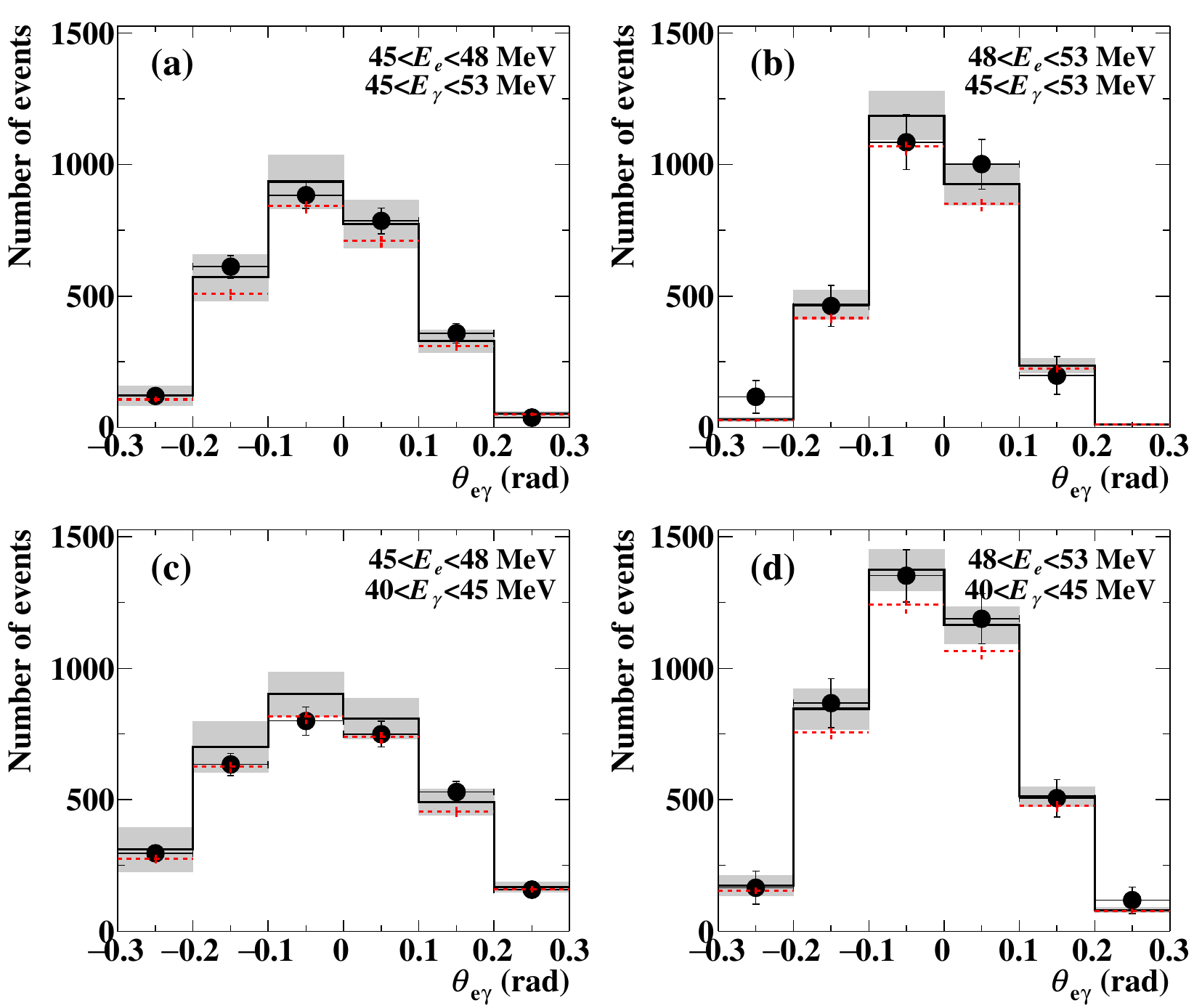}
\caption{\label{fig:RMDFit}Distribution of RMD events in the 24 bins used in the fit. 
The dots show the data and the solid lines show the expected distribution 
with $P_{\mu^+}=-0.85$ normalized by the Michel positron measurement. The bands show the systematic uncertainty 
(square root of the diagonal elements of the covariance matrix.) The red-dashed histograms 
show the best-fit distribution.}
\end{figure}

The $\chi^2$ values for $P_{\mu^+}=-0.85$ and for the best-fit value are $\chi^2(P_{\mu^+}=-0.85)/\mathrm{DOF} = 12.8 /23=0.557$ 
and $\chi^2_\mathrm{min} / \mathrm{DOF} = 11.9/22 = 0.541$, respectively, where the scale parameter $\alpha$ is at 
the best-fit value for each case. These results show consistency of the experimental spectral shape with the SM-based predictions. The distribution of measured RMD events and the calculated
ones, both for predicted parameters and for the best-fit ones, are shown in Fig.~\ref{fig:RMDFit}.

The best-fit values are $(P_{\mu^+}, \alpha) = (-0.70 \pm 0.16,0.95\pm 0.04)$. When $\alpha$ is fixed to $1$, $P_{\mu^+} = -0.71 \pm 0.15$. 
These results are consistent with Eq.~(\ref{eq:pol}) within one standard deviation.
The scale parameter $\alpha$ is weakly sensitive to the polarization, and the result changes negligibly when $P_{\mu^+}$ is fixed to $-0.85$. The result confirms the Michel normalization.

%

\section{Discussions}

\subsection{Impact on $\megsign$ search}
The measurement of RMD is a powerful internal cross-check of the experiment.
This analysis uses the same data sample, calibrations,
reconstruction and event selections as that of the $\megsign$ search in
\cite{meg2010}. 
Therefore, the agreement in the branching ratio and the distribution between the measurements and the SM predictions provides additional confidence in the reliability of the search for $\megsign$.

A more practical purpose of analysing RMD is to estimate the
number of RMD events in the fit region of the $\megsign$ search.
We extrapolate the number of RMD events measured in the low-$\egamma$
region ($40<E_\gamma<48~\mathrm{MeV}$) to the fit region ($48<E_\gamma<58~\mathrm{MeV}$) by using the ratio of the partial
branching ratios and the ratio of the efficiencies. This estimate is directly implemented 
in the likelihood of the $\megsign$ search as a constraint on the number of
RMD events \cite{meg2010,meg2013}. 

Another application of the RMD analysis is the use of RMD events as an alternative normalization channel. 
The advantage of using RMD is its closer resemblance to $\megsign$ decay compared to that of Michel decay,
since not only a positron but also a photon from a muon decay is measured. 
The Michel positron approach provides normalization with 5\% uncertainty\footnote{While 10\% uncertainty 
was assigned to the normalization in \cite{meg2009} by using the Michel channel only, 
the uncertainty was reduced to 5\% by improvements in the analysis in \cite{meg2013}; see \cite{fujii_2013} for the details.} while the RMD approach has an uncertainty of 6\%.
The systematic uncertainties of those two approaches are independent,
so that the combination leads to 4\% uncertainty in the $\megsign$ normalization \cite{meg2013}.

\subsection{RMD measurements and the future}\label{sec:future}
With the experimental precision of the RMD measurement at the level of $\sim\! 9$\%, radiative corrections to RMD are not negligible any more,
as reported in \cite{Arbuzov:2004wr,Fael:2015}, especially at the kinematic edge of its phase space where the higher order contribution could be as large as O(10\%). 
Only recently a full NLO order calculation of RMD became available \cite{Fael:2015} in addition to higher-order calculations for special cases \cite{Fischer:1994pn,Arbuzov:1998kr,Arbuzov:2004wr}.

In the next stage of $\megsign$ search experiments, such as the MEG upgrade (MEG II) \cite{Baldini:2013ke} and also in the future search for $\meeesign$ \cite{Blondel:2013ia}, detectors with higher resolutions are planned in order to reach the desired sensitivities.
They will require refined control and precise measurements of all types of background.\footnote{RMD is a source of time- and also vertex-correlated background and of accidental background for $\meeesign$ because of internal conversions of RMD photons to electron--positron pairs.} 
These measurements as well as tests of the structure of weak interactions using RMD events require precise theoretical predictions as achieved with the recent full NLO calculation which has a theoretical error well below 1.0\%. 

\section{Conclusion}
We performed the first study of radiative decay of polarized muons, $\rmdsign$.
We measured the branching ratio,
$\mathcal{B} = (6.03\pm0.14\mathrm{(stat.)}\pm0.53\mathrm{(sys.)})\times 10^{-8}$ for $\epositron>45$~MeV and $\egamma>40$~MeV, and various distributions in a large sub-sample of muon decays collected by the MEG experiment. 
Our measurement of RMD is the most precise in the kinematic region relevant to the $\megsign$ search and is consistent with the SM expectations.
The agreement with the SM strongly validates the experiment and demonstrates the capability of detecting very rare decays such as $\megsign$ in MEG.

\begin{acknowledgements}
We gratefully acknowledge the support and co-operation provided by PSI as the host laboratory and to the technical and engineering staff of our institutes.
This work is supported by DOE DEFG02-91ER40679 (USA), INFN (Italy), 
MES of Russia and RFBR Grant 14-22-03071 (Russia),
MEXT KAKENHI Grant No.~22000004, 26000004 (Japan), 
and SNF Grant No.~200021\_137738 (Switzerland). 
Partial support of MIUR Grant No. RBFR08XWGN (Italy) is acknowledged.
\end{acknowledgements}

\bibliographystyle{my}
\bibliography{./MEG}

\end{document}

%% file: author_INFNComb-epjc.tex
\newcommand*{\INFNPi}{INFN Sezione di Pisa$^{a}$; Dipartimento di Fisica$^{b}$ dell'Universit\`a, Largo B.~Pontecorvo~3, 56127 Pisa, Italy}
\newcommand*{\ScuolaPi}{Scuola Normale Superiore, Piazza dei Cvalieri 7, 56126 Pisa, Italy}
\newcommand*{\INFNGe}{INFN Sezione di Genova$^{a}$; Dipartimento di Fisica$^{b}$ dell'Universit\`a, Via Dodecaneso 33, 16146 Genova, Italy}
\newcommand*{\INFNPv}{INFN Sezione di Pavia$^{a}$; Dipartimento di Fisica$^{b}$ dell'Universit\`a, Via Bassi 6, 27100 Pavia, Italy}
\newcommand*{\INFNRm}{INFN Sezione di Roma$^{a}$; Dipartimento di Fisica$^{b}$ dell'Universit\`a ``Sapienza'', Piazzale A.~Moro, 00185 Roma, Italy}
\newcommand*{\INFNLe}{INFN Sezione di Lecce$^{a}$; Dipartimento di Matematica e Fisica$^{b}$ dell'Universit\`a del Salento, Via per Arnesano, 73100 Lecce, Italy}
\newcommand*{\ICEPP} {ICEPP, The University of Tokyo, 7-3-1 Hongo, Bunkyo-ku, Tokyo 113-0033, Japan }
\newcommand*{\UCI}   {University of California, Irvine, CA 92697, USA}
\newcommand*{\KEK}   {KEK, High Energy Accelerator Research Organization, 1-1 Oho, Tsukuba, Ibaraki 305-0801, Japan}
\newcommand*{\PSI}   {Paul Scherrer Institut PSI, 5232 Villigen, Switzerland}
\newcommand*{\Waseda}{Research Institute for Science and Engineering, Waseda~University, 3-4-1 Okubo, Shinjuku-ku, Tokyo 169-8555, Japan}
\newcommand*{\BINP}  {Budker Institute of Nuclear Physics of Siberian Branch of Russian Academy of Sciences, 630090 Novosibirsk, Russia}
\newcommand*{\JINR}  {Joint Institute for Nuclear Research, 141980 Dubna, Russia}
\newcommand*{\ETHZ}  {Swiss Federal Institute of Technology ETH, 8093 Z\" urich, Switzerland}
\newcommand*{\NOVS}  {Novosibirsk State University, 630090 Novosibirsk, Russia}
\newcommand*{\NOVST}  {Novosibirsk State Technical University, 630092 Novosibirsk, Russia}

\author{
        A.~M.~Baldini\thanksref{addINFNPi}$^{a}$ \and
        Y.~Bao\thanksref{addPSI} \and
        E.~Baracchini\thanksref{addICEPP,section} \and
        C.~Bemporad\thanksref{addINFNPi}$^{ab}$ \and
        F.~Berg\thanksref{addPSI,addETHZ} \and
        M.~Biasotti\thanksref{addINFNGe}$^{ab}$ \and 
        G.~Boca\thanksref{addINFNPv}$^{ab}$ \and
        P.~W.~Cattaneo\thanksref{addINFNPv}$^{a}$  \and
        G.~Cavoto\thanksref{addINFNRm}$^{a}$ \and
        F.~Cei\thanksref{addINFNPi}$^{ab}$ \and
        G.~Chiarello\thanksref{addINFNLe}$^{ab}$ \and
        C.~Chiri\thanksref{addINFNLe}$^{ab}$ \and
        A.~de~Bari\thanksref{addINFNPv}$^{ab}$ \and
        M.~De~Gerone\thanksref{addINFNGe}$^{a}$ \and
        A.~D\rq{}Onofrio\thanksref{addINFNPi}$^{ab}$ \and
        S.~Dussoni\thanksref{addINFNPi}$^{a}$\and
        Y.~Fujii\thanksref{addICEPP}  \and
        L.~Galli\thanksref{addINFNPi}$^{a}$ \and
        F.~Gatti\thanksref{addINFNGe}$^{ab}$ \and
        F.~Grancagnolo\thanksref{addINFNLe}$^{a}$ \and
        M.~Grassi\thanksref{addINFNPi}$^{a}$ \and
        A.~Graziosi\thanksref{addINFNRm}$^{ab}$ \and
        D.~N.~Grigoriev\thanksref{addBINP,addNOVST,addNOVS} \and
        T.~Haruyama\thanksref{addKEK} \and
        M.~Hildebrandt\thanksref{addPSI} \and
        Z.~Hodge\thanksref{addPSI,addETHZ} \and
        K.~Ieki\thanksref{addICEPP,addPSI} \and
        F.~Ignatov\thanksref{addBINP,addNOVS} \and
        T.~Iwamoto\thanksref{addICEPP}  \and
        D.~Kaneko\thanksref{addICEPP} \and
        Tae~Im~Kang\thanksref{addUCI} \and
        P.~-R.~Kettle\thanksref{addPSI} \and
        B.~I.~Khazin\thanksref{addBINP,addNOVS,dagger} \and
        N.~Khomutov\thanksref{addJINR} \and
        A.~Korenchenko\thanksref{addJINR}  \and
        N.~Kravchuk\thanksref{addJINR}  \and
        G.~M.~A.~Lim\thanksref{addUCI} \and
        S.~Mihara\thanksref{addKEK}  \and
        W.~Molzon\thanksref{addUCI} \and
        Toshinori~Mori\thanksref{addICEPP}  \and
        A.~Mtchedlishvili\thanksref{addPSI} \and
        S.~Nakaura\thanksref{addICEPP}  \and
       D.~Nicol\`o\thanksref{addINFNPi}$^{ab}$ \and
        H.~Nishiguchi\thanksref{addKEK}  \and
        M.~Nishimura\thanksref{addICEPP}  \and
        S.~Ogawa\thanksref{addICEPP}  \and
        W.~Ootani\thanksref{addICEPP}  \and
        M.~Panareo\thanksref{addINFNLe}$^{ab}$ \and
        A.~Papa\thanksref{addPSI} \and
        A.~Pepino\thanksref{addINFNLe}$^{ab}$ \and
        G.~Piredda\thanksref{addINFNRm}$^{a}$ \and
        G.~Pizzigoni\thanksref{addINFNGe}$^{ab}$ \and
        A.~Popov\thanksref{addBINP,addNOVS} \and
        F.~Renga\thanksref{addINFNRm}$^{a,}$\thanksref{addPSI} \and
        E.~Ripiccini\thanksref{addINFNRm}$^{ab}$ \and
        S.~Ritt\thanksref{addPSI} \and
        M.~Rossella\thanksref{addINFNPv}$^{a}$ \and
        G.~Rutar\thanksref{addPSI,addETHZ} \and
        R.~Sawada\thanksref{addICEPP}  \and
        F.~Sergiampietri\thanksref{addINFNPi}$^{a}$ \and
        G.~Signorelli\thanksref{addINFNPi}$^{a}$ \and
        G.~F.~Tassielli\thanksref{addINFNLe}$^{a}$ \and
        F.~Tenchini\thanksref{addINFNPi}$^{ab}$ \and
        Y.~Uchiyama\thanksref{addICEPP,e1} \and
        M.~Venturini\thanksref{addINFNPi}$^{a,}$\thanksref{addScuolaPi} \and
        C.~Voena\thanksref{addINFNRm}$^{a}$ \and
        A.~Yamamoto\thanksref{addKEK} \and
        K.~Yoshida\thanksref{addICEPP}  \and
        Z.~You\thanksref{addUCI} \and
        Yu.~V.~Yudin\thanksref{addBINP,addNOVS} 
}

\institute{\INFNPi \label{addINFNPi}
          \and	
            \ScuolaPi \label{addScuolaPi}
           \and
	\PSI \label{addPSI} 
           \and
              \ETHZ \label{addETHZ}
           \and
              \ICEPP \label{addICEPP}
          \and
             \INFNGe \label{addINFNGe}
           \and
             \INFNPv \label{addINFNPv}
          \and
             \INFNLe \label{addINFNLe}
           \and
             \INFNRm \label{addINFNRm}
          \and
             \BINP   \label{addBINP}
           \and
             \NOVST  \label{addNOVST}
           \and
             \NOVS  \label{addNOVS}
           \and
             \KEK    \label{addKEK}
           \and
             \UCI    \label{addUCI}
           \and
             \JINR   \label{addJINR}
}